\documentclass[aps,prl,twocolumn,noshowpacs,superscriptaddress]{revtex4-1}
\usepackage{graphicx,epsfig, subfigure}
\usepackage{amsbsy}
\usepackage{amssymb}
\usepackage{amsmath}
\usepackage{hyperref}
\usepackage[bottom]{footmisc}
\usepackage{color}
\usepackage{verbatim}

\date{\today}

\begin{document}

\newcommand{\eqnref}[1]{Eq.~\ref{#1}}
\newcommand{\figref}[2][]{Fig.~\ref{#2}#1}

\title{Experimental realization of Feynman's ratchet}

\author{Jaehoon Bang}
\thanks{Equal contribution}
 \affiliation{School of Electrical and Computer Engineering, Purdue University, West Lafayette, IN 47907, USA}

\author{Rui Pan}
\thanks{Equal contribution}
\affiliation{School of Physics, Peking University, Beijing 100871, China}

\author{Thai M. Hoang}
\thanks{Current address: Sandia National Laboratories, Albuquerque, NM 87123}
 \affiliation{Department of Physics and Astronomy, Purdue University, West Lafayette, IN 47907, USA}

\author{Jonghoon Ahn}
 \affiliation{School of Electrical and Computer Engineering, Purdue University, West Lafayette, IN 47907, USA}

\author{Christopher Jarzynski}
 \affiliation{Institute for Physical Science and Technology, University of Maryland, College Park, MD 20742 USA}

\author{H. T. Quan}
\email{Corresponding author: htquan@pku.edu.cn}
\affiliation{School of Physics, Peking University, Beijing 100871, China}
\affiliation{Collaborative Innovation Center of Quantum Matter, Beijing 100871, China}

\author{Tongcang Li}
 \email{Corresponding author: tcli@purdue.edu}
 \affiliation{School of Electrical and Computer Engineering, Purdue University, West Lafayette, IN 47907, USA}
  \affiliation{Department of Physics and Astronomy, Purdue University, West Lafayette, IN 47907, USA}
 \affiliation{Purdue Quantum Center, Purdue University, West Lafayette, IN 47907, USA}
 \affiliation{Birck Nanotechnology Center, Purdue University, West Lafayette, IN 47907, USA}

\begin{abstract}
{Feynman's ratchet is a microscopic machine in contact with two heat reservoirs, at temperatures $T_A$ and $T_B$, that was proposed by Richard Feynman to illustrate the second law of thermodynamics.  In equilibrium ($T_A=T_B$), thermal fluctuations prevent the ratchet from generating directed motion.  When the ratchet is maintained away from equilibrium by a temperature difference ($T_A \ne T_B$), it can operate as a heat engine, rectifying thermal fluctuations to perform work. While it has attracted much interest, the operation of Feynman's ratchet as a heat engine has not been realized experimentally, due to technical challenges. In this work, we realize Feynman's ratchet with a colloidal particle in a one dimensional optical trap in contact with two heat reservoirs: one is the surrounding water, while the effect of the other reservoir is generated by a novel feedback mechanism, using the Metropolis algorithm to impose detailed balance. We verify that the system does not produce work when $T_A=T_B$, and that it becomes a microscopic heat engine when $T_A \ne T_B$. We analyze work, heat and entropy production as functions of the temperature difference and external load. Our experimental realization of Feynman's ratchet and the Metropolis algorithm can also be used to study the thermodynamics of feedback control and information processing, the working mechanism of molecular motors, and controllable particle transportation.
}
\end{abstract}

\maketitle

In his {\it Lectures on Physics}, Richard Feynman introduced an ingenious model to illustrate the inviolability of the second law of thermodynamics \cite{Feynman2006}. A ratchet and pawl are arranged to permit a wheel to turn in only one direction, and the wheel is attached to a windmill immersed in a gas.  Random collisions of gas molecules against the windmill's panes would then seemingly drive the wheel to rotate systematically in the allowed direction. This  rotation could be used to extract useful work from the thermal fluctuations in the gas, in flagrant violation of the second law of thermodynamics. As discussed by Feynman, this violation does not occur because thermal fluctuations of the pawl occasionally allow the ratchet to move in the ``forbidden'' direction. However, if the pawl is maintained at a temperature that differs from that of the gas, then the device is indeed able to rectify thermal fluctuations to produce work -- in this case it operates as a microscopic heat engine.

\begin{figure}[b!]
	\includegraphics[scale=0.48]{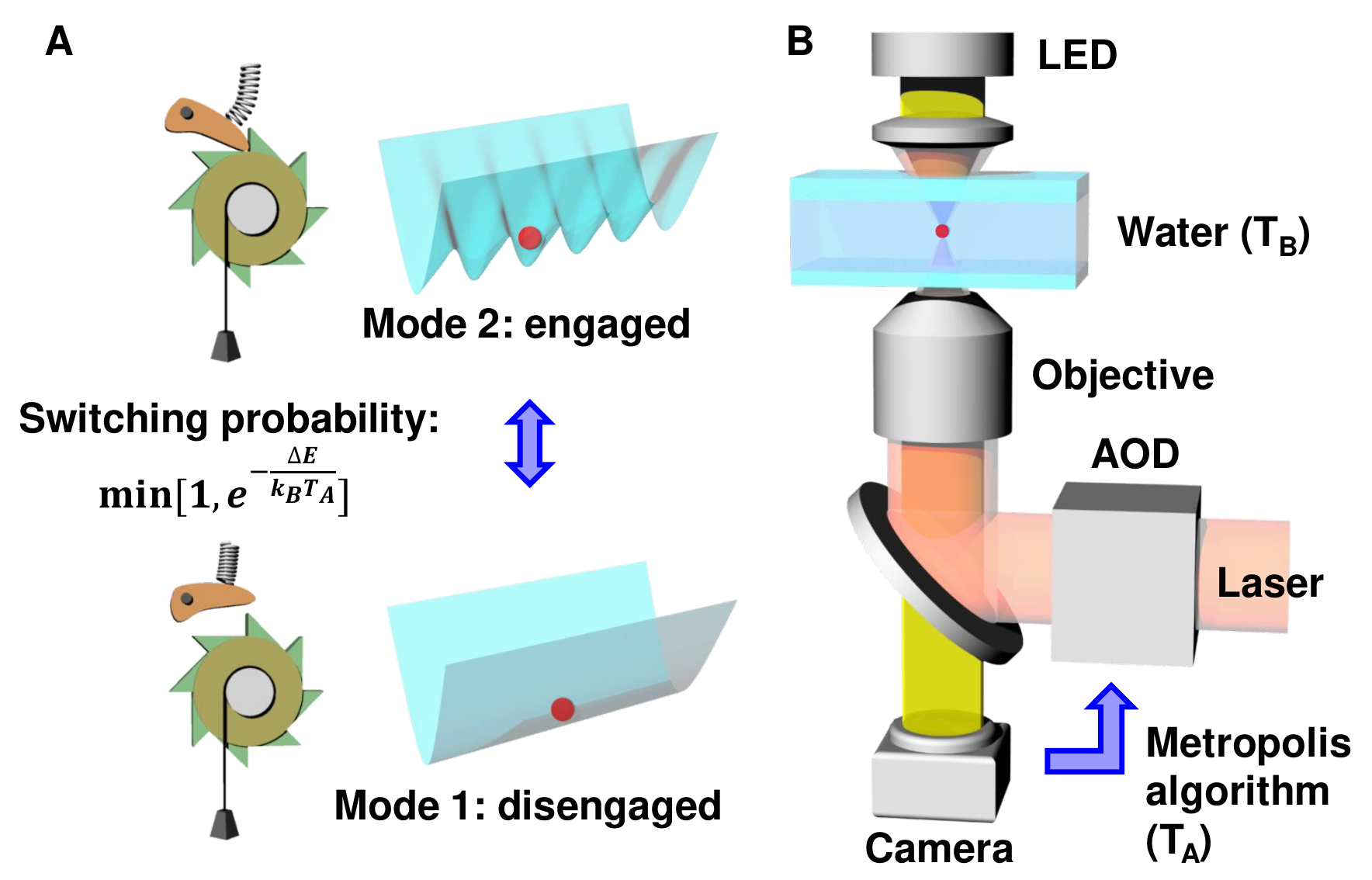}
	\caption{Schematic drawings of the experiment. A, A sawtooth potential and a uniform potential correspond to the engaged and disengaged modes, respectively, of the ratchet and pawl. We switch between the two potential modes following the Metropolis algorithm to generate the effects of a heat reservoir at temperature $T_A$. B, A 780-nm-diameter silica microsphere is trapped in a 1D optical trap inside a water chamber using a 1064 nm laser. The water is at temperature $T_B=296$ K. The location of the particle is recorded by a camera for computer feedback control. AOD: acousto-optic deflector.
	}
	\label{Fig:schematic}
\end{figure}

Three essential features are needed to produce directed motion in Feynman's model: 1) when the ratchet and pawl are engaged, the potential energy profile must be asymmetric; 2) the device must be in contact with thermal reservoirs at different temperatures; and 3) the device must be  small enough to undergo Brownian motion driven by thermal noise.  Extensive theoretical analyses of Feynman's device \cite{Magnasco1993,Parrondo1996,Sekimoto1997,Astumian1998,Magnasco1998,Jarzynski1999,VandenBroeck2004,Zheng2010} and related ratchet models \cite{Reimann2002} have been used to gain insight into motor proteins \cite{Astumian1994, Prost1994, Astumian1996, Kolomeisky1998}. Simpler models with a single heat reservoir have been proposed, and experimental demonstration of the directed Brownian motion in asymmetric potentials had been accomplished \cite{Rousselet1994,Faucheux1995,Matthias2003,Lopez2008,Wu2016,Arzola2017}.  Recently, a macroscopic (10-cm-scale) ratchet driven by mechanical collisions of 4-mm-diameter glass beads was demonstrated \cite{Eshuis2010}.  However, an experimental realization of Feynman's ratchet, which can rectify thermal fluctuations from two heat reservoirs to perform work, has not been reported to date.
A major challenge is to devise a microscopic system that is in contact with two heat reservoirs at different temperatures, without side effects such as fluid convection that can smear the effects of thermal fluctuations.

Here we realize Feynman's two-temperature ratchet and pawl with a colloidal particle confined in a one-dimensional (1D) optical trap (Fig. \ref{Fig:schematic}).
The particle's 1D Brownian motion emulates the collision-driven rotation of the ratchet, and we use optical tweezers to generate both uniform and sawtooth potentials, simulating respectively the disengaged and engaged modes of the pawl (Fig. \ref{Fig:schematic}A).
The uniform potential is smooth enough that the colloidal particle moves freely in the disengaged mode.
The water surrounding the colloidal particle provides a heat reservoir at temperature $T_B$.
The other heat reservoir is generated by using feedback control of the optical tweezer array \cite{Endres2016} to toggle between the disengaged and engaged modes of the pawl, implementing the Metropolis algorithm \cite{Metropolis1953} as in Ref. \cite{Jarzynski1999} to satisfy the detailed balance condition at a chosen temperature $T_A$ (Eq. (\ref{eq:switch})).
Our setup, inspired by both continuous \cite{Parrondo1996,Sekimoto1997,Magnasco1998} and discrete \cite{Jarzynski1999} theoretical models, captures the essential features of Feynman's original model, with the pawl maintained at one temperature ($T_A$) and the ratchet at another ($T_B$). As described in detail below, both numerical simulation and experimental data clearly show the influence of the two heat reservoirs on the operation of the ratchet, in agreement with theoretical analyses \cite{Feynman2006,Jarzynski1999}.

In our experiment, a silica microsphere with a diameter of 780 nm, immersed in deionized water, undergoes diffusion in a 1D optical trap created with an array of 19 optical tweezers. The array is generated by an acousto-optic deflector (AOD) controlled by an arbitrary function generator \cite{Endres2016}. By  tuning the power of each optical tweezer individually, the trap potential can be modified to be either  uniform or sawtooth-shaped.
The water chamber mimics the gas heat reservoir in Feynman's model; its temperature $T_B$ was measured to be around 296 K.
We use computer feedback to generate the effects of an artificial heat reservoir, at temperature $T_A$, coupled to the pawl as it switches between its two modes: (1) disengaged and (2) engaged.
Letting $U_1(x)$ and $U_2(x)$ denote the potential energies of these modes, as functions of the particle location $x$, we generate attempted switches between modes at a rate $\Gamma$, and each such attempt is accepted with a probability given by the Metropolis algorithm \cite{Metropolis1953,Jarzynski1999}:
\begin{eqnarray}
P_{\rm switch}=\min[1,\exp(-\Delta E/k_BT_A)],
\label{eq:switch}
\end{eqnarray}
where $k_B$ is Boltzmann's constant, and $\Delta E = U_j(x)-U_i(x)$ for an attempted switch from mode $i$ to mode $j$.

\begin{figure}[t!]
	\includegraphics[scale=0.48]{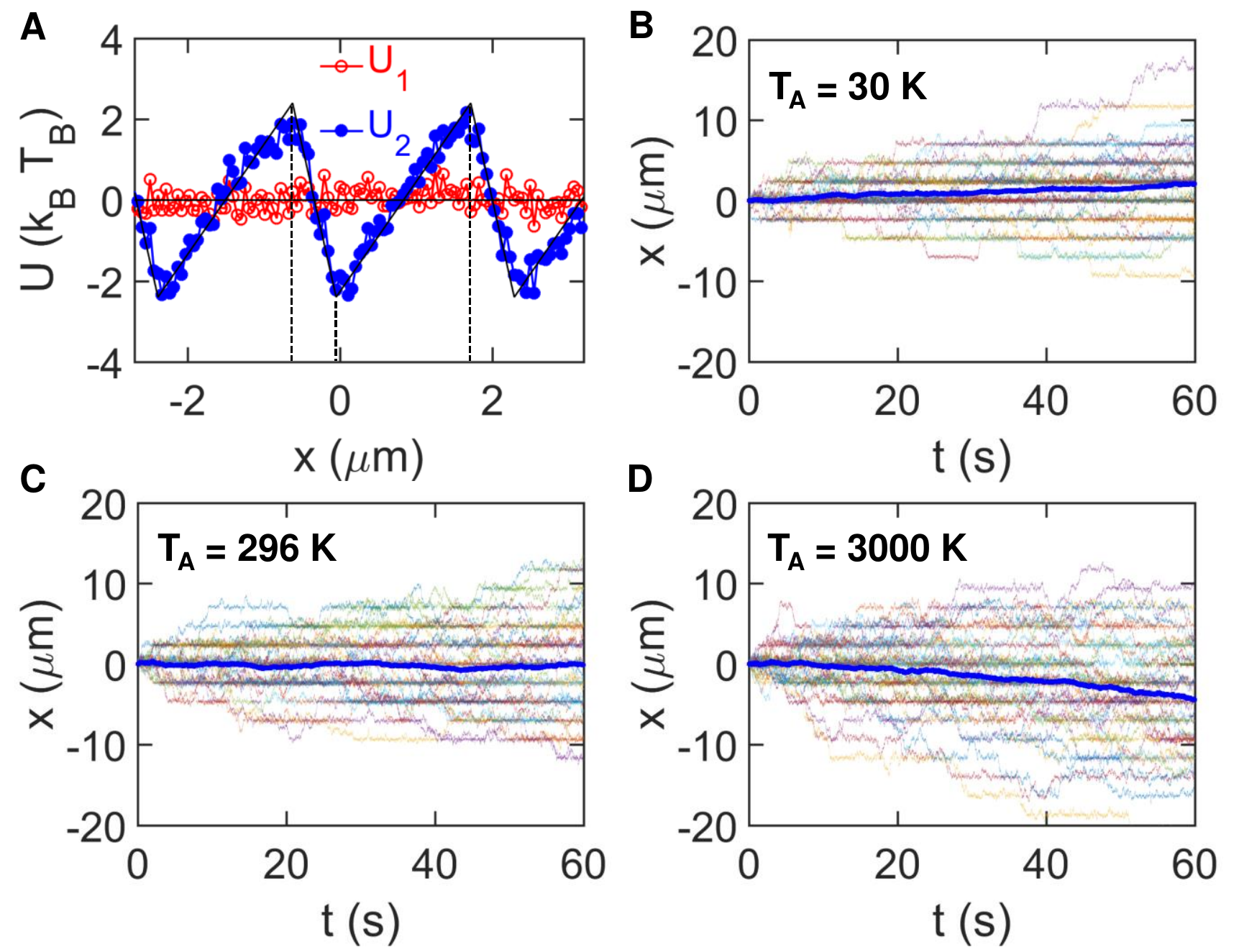}
	\caption{Potential profiles of the 1D optical trap and trajectories of the microsphere. A, Measured mode 1 (red) and mode 2 (blue) potential profiles at room temperature $T_B = 296$ K. Black lines indicate the fitted linear potentials used in numerical simulations. The asymmetry in mode 2 is about 1:3. B-D, 50 individual trajectories (thin lines) and the average trajectory (thick blue line) of the microsphere when the second heat reservoir is at temperature $T_A = 30$ K (B), $T_A = 296$ K (C), and  $T_A = 3000$ K (D). The average displacement of the particle at 60 s is $\left\langle \Delta x \right\rangle =2.1 \pm 0.6  \mu$m in B, $\left\langle \Delta x \right\rangle $=$-0.1 \pm 0.9 \mu$m in C, and $\left\langle \Delta x \right\rangle $=$-4.5 \pm 0.9 \mu$m in D.
	}
	\label{Fig:trajectories}
\end{figure}

To model the dynamics of our system, we couple the diffusive motion of the Brownian particle in the 1D trap to the stochastic switching between the engaged and disengaged potential modes.
Letting $P_i(x,t)$ denote the joint probability density to find the particle at position $x$ and the potential in mode $i \in \{1,2\}$ at time $t$, we construct the reaction-diffusion equation \cite{Astumian1998}
\begin{eqnarray}
\frac{\partial P_i(x,t)}{\partial t}=&&\frac{\partial}{\partial x} \left[\frac{U_i'(x)}{\gamma} P_i (x,t)\right]+D \frac{\partial^2 P_i (x,t)}{\partial x^2} \nonumber\\
&&+k_{ji}(x) P_j (x,t)-k_{ij}(x) P_i (x,t),
	\label{reactiondiffusion}
\end{eqnarray}
where $\gamma$ is the Stokes friction coefficient, $D=k_B T_B/\gamma$ the diffusion constant, and $k_{ij}$ ($k_{ji}$)($i \neq j$) the switching rate from potential mode $i(j)$ to $j(i)$.
This rate is the product of the constant attempt rate, $\Gamma$, and the Metropolis acceptance probability, Eq. (\ref{eq:switch}):
\begin{eqnarray}
k_{ij}=\Gamma \min \left[1, \exp \left(-\frac{U_j (x)-U_i (x)}{k_B T_A} \right)\right],\label{metropolisalgorithm1}
\end{eqnarray}
In our experiment, $\Gamma$ is chosen to optimize the particle velocity.
The total particle probability distribution is $P_{1}(x,t)+P_{2}(x,t)$.

The  potentials $U_1(x)$ and $U_2(x)$ are determined from the measured equilibrium distribution of the particle in each mode. We fix the 1D trap in a particular mode and we track the Brownian motion of the silica microsphere in the trap, recording its position every 5~ms. After more than $5\times 10^5$ data points are collected, the potential profile is extracted using the equation $U(x)= -k_B T_B \ln[N(x)/N_{total}]$, where $N(x)$ is the number of count at each point and $N_{total}$ is the total count.  The position $x$ is discretized in bins of 52~nm corresponding to the pixel size of our camera. The measured uniform and sawtooth potentials are shown in Fig. \ref{Fig:trajectories}A. The depth (from trough to peak) of the sawtooth potential $U_2(x)$ is about 4.8 $k_B T_B$. The sawtooth potential is described by an asymmetry ratio of approximately 1:3 (see Fig.~\ref{Fig:trajectories}A); this is a key parameter for thermal ratchets \cite{Hanggi2009}. The uniform potential $U_1(x)$ has a standard deviation of about 0.15 $k_B T_B$, which is small enough for the silica microsphere to diffuse freely.  After measuring the potentials, we fit $U_1(x)$ with a straight line and $U_2(x)$  with an ideal sawtooth potential (Fig. \ref{Fig:trajectories}A). These fitted smooth potentials were used in our numerical simulations to avoid the statistical noise in the measurements.

To implement the ratchet dynamics, we first trap a silica microsphere  with a single optical tweezer and position it near the middle of the trap, which corresponds to the potential minimum at the center of mode 2. The optical tweezer array is then turned on at mode 2, and we follow the diffusion of the microsphere as the potential switches between modes as described earlier. The position of the microsphere is recorded every 5 ms using a complementary metal-oxide semiconductor (CMOS) camera (Fig. \ref{Fig:schematic}B). Every 200 ms (this time interval is equal to $1/\Gamma$), an attempt to switch modes is made, and is accepted with the probability given by Eq.~(\ref{eq:switch}). In our experiment, the sawtooth potential has finite length. To mimic the infinite nature of a rotating ratchet, the particle is dragged back to the trap center whenever it reaches one of the potential minima located on both ends.

\begin{figure}[t!]
	\includegraphics[scale=0.9]{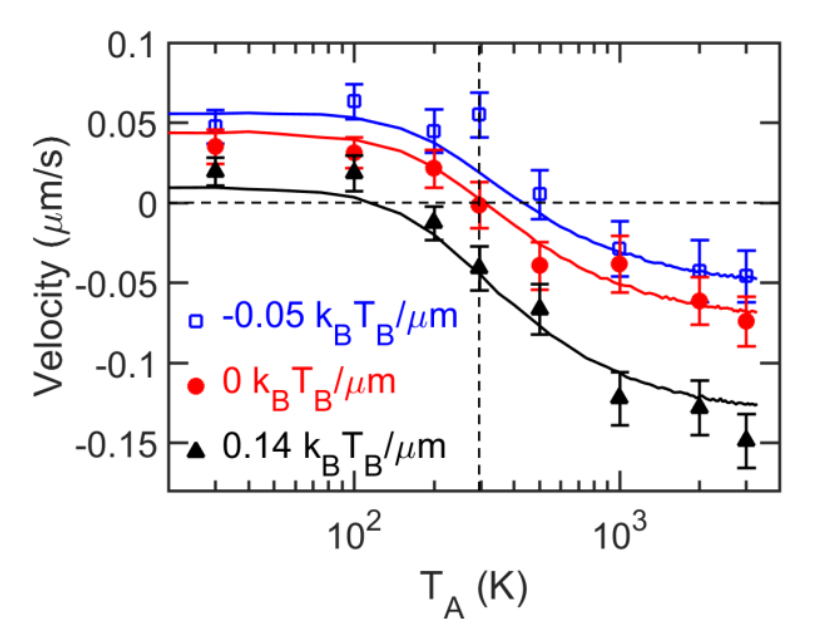}
	\caption{Average velocity of the microsphere under different external loads  at different temperatures $T_A$. Blue squares, red circles and black triangles represent results with potential slopes of $-0.05$, $0$ and $0.14$ $k_B T_B/\mu$m respectively. The added slope to the original potential represents the external load. Every experimental data point represents the result of an average of 50 repetitions. Solid lines are simulation results. The vertical dashed line indicates $T_B=296$~K.
	}
	\label{Fig:velocity}
\end{figure}

Figs.~\ref{Fig:trajectories}B, \ref{Fig:trajectories}C, and \ref{Fig:trajectories}D show 60-s trajectories of the particle  without external load for different heat reservoir temperatures $T_A$.  The light lines display individual trajectories, and the thick blue lines are averages over these trajectories.
We interpret positive displacements of the particle as clockwise rotation of the mechanical ratchet, and negative displacements as counter-clockwise rotation.
As seen in Fig. \ref{Fig:trajectories}C,  the average final displacement of the particle converges essentially to zero,  $\left\langle \Delta x \right\rangle =-0.1 \pm 0.9 \mu$m, when the temperatures  of the two reservoirs are equal ($T_A=T_B=296$K)  even though the potential is asymmetric. This experimentally verifies Feynman's prediction that the ratchet does not produce perpetual motion, as clockwise rotations are cancelled by counter-clockwise rotations, on average.
When $T_A=30$K ($<T_B$), the average final position of the microsphere is $\left\langle \Delta x \right\rangle =2.1 \pm 0.6  \mu$m, indicating net clockwise rotation of the ratchet.
When $T_A=3000$K ($>T_B$), the average final position is observed to be $\left\langle \Delta x \right\rangle =-4.5 \pm 0.9  \mu$m, corresponding to net counter-clockwise rotation.
These results demonstrate that a temperature difference $T_B-T_A$ can give rise to unidirectional motion via the rectification of thermal noise \cite{Astumian1998}.
We now investigate whether this motion can be harnesses to perform work against an external load.

\begin{figure*}[t!]
	\includegraphics[scale=0.6]{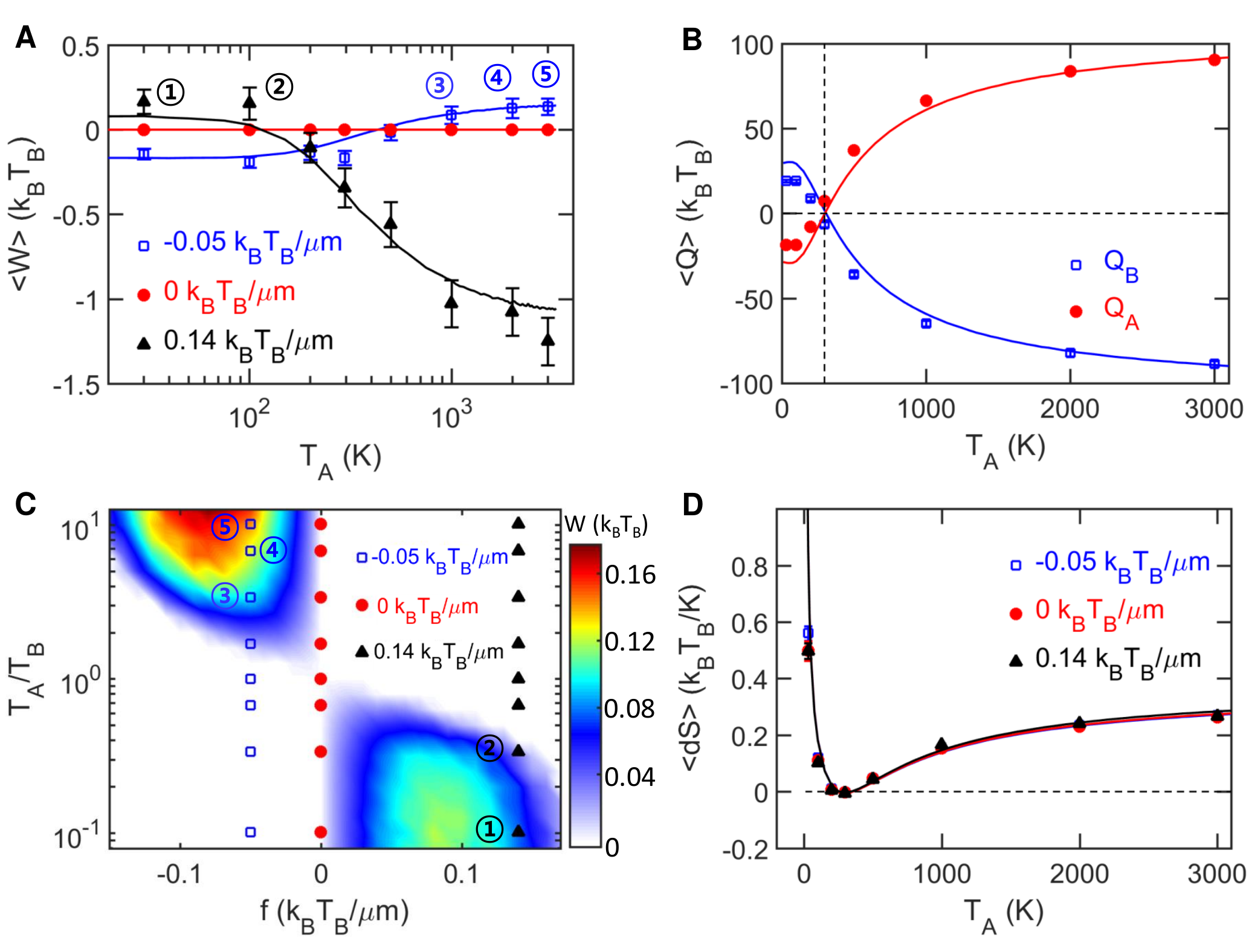}
	\caption{Work, heat, and entropy production of Feynman's ratchet. A, Extracted work values when the added potential slope is $-0.05$ $k_B T_B/\mu$m (blue squares), $0$ $k_B T_B/\mu$m (red circles) and $0.14$ $k_BT_B/\mu$m (black triangles). Error bars show the uncertainty of the average. B, Heat absorbed from reservoir A (red circles) and B (blue squares) in the presence of $-0.05$ $k_BT_B/\mu m$ slope is calculated for each trajectory. Each data point shows the average heat calculated from 50 experimental trajectories. Error bars indicating the uncertainty are smaller than the symbols. Solid lines are simulation results. Dashed vertical line indicates $T_{B}=296K$. C, The numerically calculated work map of Feynman's ratchet with representative points corresponding to Fig. 3 and Fig. 4A. The color shaded area is the heat engine regime where the simulated work is positive.  D, average entropy production in 60~s as a function of $T_A$.
	}
	\label{Fig:work}
\end{figure*}

To apply an effective external load to the ratchet, a slight linear slope $f$ is added to the potentials. Fig. \ref{Fig:velocity} shows the observed average particle velocity as a function of pawl temperature $T_A$, for $f=-0.05 k_BT_B/\mu$m, $f=0$ and $f=0.14 k_BT_B/\mu$m, corresponding to positive, zero and negative external loads, respectively.
Each experimental data point is calculated from fifty 60-s trajectories.
The lines show numerical simulation results, with each situation simulated over $5 \times 10^4$ times to achieve high accuracy.
In these simulations we use the overdamped Langevin equation $\dot{x}=-\frac{1}{\gamma}  \frac{\partial U}{\partial x}+\xi (t)$, where $\xi (t)$ is Gaussian random force satisfying $\left\langle \xi (t) \right\rangle =0$, $\left\langle \xi (t)\xi(t') \right\rangle =\frac{2k_B T_B}{\gamma} \delta(t-t')$, and $\gamma$ is the Stokes friction coefficient. The simulation time step is 2~ms. The simulations are performed with the same procedure as the experiment, described earlier.
The simulation and experimental results show good agreement over a wide range of $T_A$ (Fig. \ref{Fig:velocity}).

By realizing an external load, we can interpret our system as a microscopic heat engine. The work produced by the engine is given by the product of the slope and the final displacement of the particle $W =f \Delta x$.  In the case of a positive external load ($f<0$), data points in Fig.~\ref{Fig:velocity} with negative average velocity correspond to positive average work performed by the engine. Conversely, in the negative loading case ($f>0$), data points in Fig.~\ref{Fig:velocity} having positive average velocity correspond to positive average work performed by the engine.

The thermodynamic operation of our system as a heat engine is further illustrated in Fig. \ref{Fig:work}.
In Fig. \ref{Fig:work}A we plot the average work as a function of $T_A$, with the points labeled by circled integers indicating the parameters for which the system generates positive work.
For our parameter choices, the greatest observed amount of work extracted in 60~s is 0.14 $k_B T_B$ when $f=-0.05 k_B T_B/ \mu$m, and 0.16 $k_BT_B$ when $f=0.14 k_BT_B/\mu m$.
Recall that in the absence of external load, our system generates positive velocities when $T_A<T_B$ and negative ones when $T_A>T_B$ (Fig. \ref{Fig:trajectories}), suggesting that in these situations it might be able to perform work when $f>0$ and $f<0$, respectively.
The points corresponding to $\left\langle W\right\rangle >0$ in Fig. \ref{Fig:work}A confirm this expectation. (There is a minor exception when $f=-0.05 k_B T_B/ \mu$m and $T_A=500$K, where the experimental uncertainty is too large to observe the very small predicted positive work.)
Similarly, in the phase diagram shown in Fig. \ref{Fig:work}C, we see that the system acts as a heat engine (i.e. generates positive work) only when the sign of $T_A-T_B$ is opposite to the sign of $f$. 

Changes in the potential energy of the particle due to diffusion are associated with heat exchange with reservoir B, and changes in its potential energy during switches between the two modes are associated with heat exchange with reservoir A.
For each trajectory, we can thus calculate the net heat absorbed by the system from the two reservoirs, $Q_B$ and $Q_A$.
For the case of positive load, Fig. \ref{Fig:work}B plots the average values of these quantities.
Note that these data show that the system absorbs energy ($Q>0$) from the hotter reservoir and releases energy ($Q<0$) into the colder reservoir, in agreement with expectations.
We have also computed the average entropy production $\left\langle dS \right\rangle=-\frac{\left\langle Q_{A}\right\rangle}{T_{A}}-\frac{\left\langle Q_{B}\right\rangle}{T_{B}}$, shown in Fig. \ref{Fig:work}D.
As expected, $\left\langle dS \right\rangle=0$ when $T_A=T_B$ as the system is then in equilibrium, but $\left\langle dS \right\rangle>0$ when $T_A\ne T_B$, in agreement with the second law.
In our model, most of the energy absorbed from the hot reservoir is delivered to the cold reservoir, with only a very small portion converted to work -- this explains the observation that the entropy production is largely independent of external load, as seen in Fig. \ref{Fig:work}D.


Fig. \ref{Fig:work} shows agreement between experiments (points) and simulations (lines), and demonstrates the operation of Feynman's ratchet as an engine that rectifies thermal fluctuations to perform work.  It is interesting to consider the thermodynamic efficiency of the engine, $\eta=\frac{\left\langle W \right\rangle}{\left\langle Q_{high} \right\rangle}$, where $Q_{high}$ denotes the heat from the reservoir with a higher temperature.
Although Feynman suggested that his ratchet could achieve Carnot efficiency \cite{Feynman2006}, later authors argued that his analysis was incorrect \cite{Parrondo1996,Sekimoto1997,Magnasco1998}.
The efficiency that we measured experimentally is $\eta=0.0015$ for the data point labeled by a circled ``5'' in Fig. \ref{Fig:work}A, which is much lower than the corresponding Carnot efficiency $\eta_C = 0.9$, in agreement with the conclusions of Refs. \cite{Parrondo1996,Sekimoto1997,Magnasco1998}.
Feynman's ratchet cannot achieve Carnot efficiency, as it generates work by continuously rectifying fluctuations in a nonequilibrium steady state \cite{Parrondo1996} -- this mode of operation is fundamentally different from the thermodynamic cycle of reversible expansion and compression that characterizes a Carnot engine.
We note that micron-sized heat engines that operate in cycles have recently been implemented in experiments using colloidal particles \cite{Blickle2012,martinez2016,Krishnamurthy2016}.
We speculate that our system can be modified to improve its efficiency by triggering the  Metropolis algorithm only when the Brownian particle reaches specific locations.

In conclusion, by combining the Metropolis algorithm with feedback control, we have realized Feynman's two-temperature ratchet-and-pawl model using a silica microsphere confined in a computer controlled one dimensional optical trap. When the temperatures of the two heat reservoirs are equal, we find no unidirectional average drift of the particle, in agreement with the predictions of Feynman and others, and with the second law of thermodynamics. When the temperatures differ, we demonstrate that thermal fluctuations can be rectified by the ratchet to generate work.  Our system provides a versatile testbed for studying the nonequilibrium thermodynamics of microscopic heat engines and molecular motors \cite{Astumian1994,Astumian1996,Kolomeisky1998,Toyabe2015,Prost1997}.  For instance, multiple heat reservoirs can be mimicked by using a position-dependent $T_A(x)$ in the Metropolis algorithm.
Moreover, although we have used feedback control to implement an effective heat bath, in an alternative scenario feedback control could be used to mimic the operation of Maxwell's demon \cite{Maxwell2001}, and thus to investigate issues related to the thermodynamics of information processing \cite{Maruyama2009,Sagawa2012,Parrondo2015}.
Our study may also have potential applications in particle transportation \cite{Astumian1998} and separation \cite{Bader1999} induced by Brownian motion in asymmetric potential.






\end{document}